%%%%%%%%%%%%%%%%%%%%%%%%%%%%%%%%%%%%%%%%%%%%%%%%%%%%%%%%%%%%%%%%%%%%%
%%% This paper is written in plain Tex and requires no input macros.
%%% It already includes a shortened version of jnl.tex and reforder.tex
     
\font\twelverm=cmr10 scaled 1200    \font\twelvei=cmmi10 scaled 1200
\font\twelvesy=cmsy10 scaled 1200   \font\twelveex=cmex10 scaled 1200
\font\twelvebf=cmbx10 scaled 1200   \font\twelvesl=cmsl10 scaled 1200
\font\twelvett=cmtt10 scaled 1200   \font\twelveit=cmti10 scaled 1200
\font\twelvesc=cmcsc10 scaled 1200  %\font\twelvesf=cmssmc10 scaled 1200
\skewchar\twelvei='177   \skewchar\twelvesy='60
     
%  Define \...point macros to change fonts and spacings consistently
     
\def\twelvepoint{\normalbaselineskip=12.4pt plus 0.1pt minus 0.1pt
  \abovedisplayskip 12.4pt plus 3pt minus 9pt
  \belowdisplayskip 12.4pt plus 3pt minus 9pt
  \abovedisplayshortskip 0pt plus 3pt
  \belowdisplayshortskip 7.2pt plus 3pt minus 4pt
  \smallskipamount=3.6pt plus1.2pt minus1.2pt
  \medskipamount=7.2pt plus2.4pt minus2.4pt
  \bigskipamount=14.4pt plus4.8pt minus4.8pt
  \def\rm{\fam0\twelverm}          \def\it{\fam\itfam\twelveit}%
  \def\sl{\fam\slfam\twelvesl}     \def\bf{\fam\bffam\twelvebf}%
  \def\mit{\fam 1}                 \def\cal{\fam 2}%
  \def\sc{\twelvesc}               \def\tt{\twelvett}
  \def\sf{\twelvesf}
  \textfont0=\twelverm   \scriptfont0=\tenrm   \scriptscriptfont0=\sevenrm
  \textfont1=\twelvei    \scriptfont1=\teni    \scriptscriptfont1=\seveni
  \textfont2=\twelvesy   \scriptfont2=\tensy   \scriptscriptfont2=\sevensy
  \textfont3=\twelveex   \scriptfont3=\twelveex  \scriptscriptfont3=\twelveex
  \textfont\itfam=\twelveit
  \textfont\slfam=\twelvesl
  \textfont\bffam=\twelvebf \scriptfont\bffam=\tenbf
  \scriptscriptfont\bffam=\sevenbf
  \normalbaselines\rm}
     
%       tenpoint

%%
%%      Various internal macros
%%
     
\def\beginlinemode{\endmode
  \begingroup\parskip=0pt \obeylines\def\\{\par}\def\endmode{\par\endgroup}}
\def\beginparmode{\endmode
  \begingroup \def\endmode{\par\endgroup}}
\let\endmode=\par
{\obeylines\gdef\
{}}
\def\singlespace{\baselineskip=\normalbaselineskip}

\def\oneandahalfspace{\baselineskip=\normalbaselineskip
  \multiply\baselineskip by 3 \divide\baselineskip by 2}
\def\doublespace{\baselineskip=\normalbaselineskip \multiply\baselineskip by 2}

\newcount\firstpageno
\firstpageno=2
\footline={\ifnum\pageno<\firstpageno{\hfil}\else{\hfil\twelverm\folio\hfil}\fi}
\def\toppageno{\global\footline={\hfil}\global\headline
  ={\ifnum\pageno<\firstpageno{\hfil}\else{\hfil\twelverm\folio\hfil}\fi}}
\let\rawfootnote=\footnote              % We must set the footnote style
\def\footnote#1#2{{\rm\singlespace\parindent=0pt\parskip=0pt
  \rawfootnote{#1}{#2\hfill\vrule height 0pt depth 6pt width 0pt}}}
\def\raggedcenter{\leftskip=4em plus 12em \rightskip=\leftskip
  \parindent=0pt \parfillskip=0pt \spaceskip=.3333em \xspaceskip=.5em
  \pretolerance=9999 \tolerance=9999
  \hyphenpenalty=9999 \exhyphenpenalty=9999 }
\def\dateline{\rightline{\ifcase\month\or
  January\or February\or March\or April\or May\or June\or
  July\or August\or September\or October\or November\or December\fi
  \space\number\year}}
\def\received{\vskip 3pt plus 0.2fill
 \centerline{\sl (Received\space\ifcase\month\or
  January\or February\or March\or April\or May\or June\or
  July\or August\or September\or October\or November\or December\fi
  \qquad, \number\year)}}
     
%%
%%      Page layout, margins, font and spacing (feel free to change)
%%
     
\hsize=6.5truein
%\hoffset=1truein
\vsize=8.5truein
%\voffset=-1.0truein
\parskip=\medskipamount
\def\\{\cr}
\twelvepoint            % selects twelvepoint fonts (cf. \tenpoint)
\doublespace            % selects double spacing for main part of paper (cf.
                        %       \singlespace, \oneandahalfspace)
\overfullrule=0pt       % delete the nasty little black boxes for overfull box

\def\title                      %  Title on title page
  {\null\vskip 3pt plus 0.2fill
   \beginlinemode \doublespace \raggedcenter \bf}
     
\def\author                     %  Author(s) name(s)  on title page
  {\vskip 3pt plus 0.2fill \beginlinemode
   \singlespace \raggedcenter\sc}
     
\def\affil                      % Affiliations (can intermix with \author)
  {\vskip 3pt plus 0.1fill \beginlinemode
   \oneandahalfspace \raggedcenter \sl}
     
\def\abstract                   % Begin abstract
  {\vskip 3pt plus 0.3fill \beginparmode
   \singlespace ABSTRACT: }
     
\def\endtopmatter               % End title page, begin body of paper
  {\endpage                     %       This subsumes \body
   \body}
     
\def\body                       % Begin text body;  can be used to end
  {\beginparmode}               % \title, \author, \affil, \abstract,
                                % \reference, or \figurecaption modes
     
\def\head#1{                    % Head;  NOTE enclose the text in {}
  \goodbreak\vskip 0.5truein    %  e.g., \head{I. Introduction}
  {\immediate\write16{#1}
   \raggedcenter \uppercase{#1}\par}
   \nobreak\vskip 0.25truein\nobreak}

\def\beginitems{
\par\medskip\bgroup\def\i##1 {\item{##1}}\def\ii##1 {\itemitem{##1}}
\leftskip=36pt\parskip=0pt}
\def\enditems{\par\egroup}
     
\def\beneathrel#1\under#2{\mathrel{\mathop{#2}\limits_{#1}}}
     
\def\refto#1{$^{#1}$}           % For references in text as superscript
     
\def\references                 % Begin references -- basic format is Phys Rev
  {\head{References}            % I.e., volume, page, year (space after commas).
   \beginparmode
   \frenchspacing \parindent=0pt \leftskip=1truecm
   \parskip=8pt plus 3pt \everypar{\hangindent=\parindent}}

\gdef\refis#1{\item{#1.\ }}                     % Ref list numbers.
     
\gdef\journal#1, #2, #3, 1#4#5#6{               % Journal reference.  Comma sets
    {\sl #1~}{\bf #2}, #3 (1#4#5#6)}            % off: name, vol, page, year

\gdef\refa#1, #2, #3, #4, 1#5#6#7.{\noindent#1, #2 {\bf #3}, #4 (1#5#6#7).\rm} 
%refa: type in: name, 
%journal, vol, page, year
%prints out in same order

\gdef\refb#1, #2, #3, #4, 1#5#6#7.{\noindent#1 (1#5#6#7), #2 {\bf #3}, #4.\rm} 
%refb: reads in same
%prints out name (year) etc.

\def\pr{\journal Phys.Rev., }

\def\prl{\journal Phys.Rev.Lett., }

\def\cmp{\journal Comm.Math.Phys., }
     
\def\np{\journal Nucl.Phys., }

\def\annp{\journal Ann.Phys.(N.Y.), }

\def\endreferences{\body}

\def\endpage                    %  Eject a page
  {\vfill\eject}
     
\def\endpaper                   %  Ways to say goodbye
  {\endmode\vfill\supereject}

\def\ref#1{Ref.~#1}                     %       for inline references
\def\Ref#1{Ref.~#1}                     %       ditto
\def\[#1]{[\cite{#1}]}
\def\cite#1{{#1}}
\def\(#1){(\call{#1})}
\def\call#1{{#1}}
\def\taghead#1{}
\def\frac#1#2{{#1 \over #2}}
\def\half{{\frac 12}}

\def\12{{1\over2}}

\catcode`@=11
\newcount\r@fcount \r@fcount=0
\newcount\r@fcurr
\immediate\newwrite\reffile
\newif\ifr@ffile\r@ffilefalse
\def\w@rnwrite#1{\ifr@ffile\immediate\write\reffile{#1}\fi\message{#1}}

\def\writer@f#1>>{}
\def\referencefile{%			  Stuff to write .REF file
  \r@ffiletrue\immediate\openout\reffile=\jobname.ref%
  \def\writer@f##1>>{\ifr@ffile\immediate\write\reffile%
    {\noexpand\refis{##1} = \csname r@fnum##1\endcsname = %
     \expandafter\expandafter\expandafter\strip@t\expandafter%
     \meaning\csname r@ftext\csname r@fnum##1\endcsname\endcsname}\fi}%
  \def\strip@t##1>>{}}

\def\citeall#1{\xdef#1##1{#1{\noexpand\cite{##1}}}}
\def\cite#1{\each@rg\citer@nge{#1}}	% Variable No. of args, separated by

\def\each@rg#1#2{{\let\thecsname=#1\expandafter\first@rg#2,\end,}}
\def\first@rg#1,{\thecsname{#1}\apply@rg}	% each@ag is a general purpose
\def\apply@rg#1,{\ifx\end#1\let\next=\relax%	  variable no. of arg. macro.
\else,\thecsname{#1}\let\next=\apply@rg\fi\next}% args separated by commas

\def\citer@nge#1{\citedor@nge#1-\end-}	% Check for M-N range (M and N numbers)
\def\citer@ngeat#1\end-{#1}
\def\citedor@nge#1-#2-{\ifx\end#2\r@featspace#1 % Single argument
  \else\citel@@p{#1}{#2}\citer@ngeat\fi}	% M-N range of arguments
\def\citel@@p#1#2{\ifnum#1>#2{\errmessage{Reference range #1-#2\space is bad.}%
    \errhelp{If you cite a series of references by the notation M-N, then M and
    N must be integers, and N must be greater than or equal to M.}}\else%
 {\count0=#1\count1=#2\advance\count1 by1\relax\expandafter\r@fcite\the\count0,
  \loop\advance\count0 by1\relax%	  Loop from M to N
    \ifnum\count0<\count1,\expandafter\r@fcite\the\count0,%
  \repeat}\fi}

\def\r@featspace#1#2 {\r@fcite#1#2,}	% Eat spaces at beginning or end of arg
\def\r@fcite#1,{\ifuncit@d{#1}%		  Cite individual reference
    \newr@f{#1}%
    \expandafter\gdef\csname r@ftext\number\r@fcount\endcsname%
                     {\message{Reference #1 to be supplied.}%
                      \writer@f#1>>#1 to be supplied.\par}%
 \fi%
 \csname r@fnum#1\endcsname}
\def\ifuncit@d#1{\expandafter\ifx\csname r@fnum#1\endcsname\relax}%
\def\newr@f#1{\global\advance\r@fcount by1%
    \expandafter\xdef\csname r@fnum#1\endcsname{\number\r@fcount}}

\let\r@fis=\refis			% Save old \refis, redefine
\def\refis#1#2#3\par{\ifuncit@d{#1}%      Use two params #2 #3 to strip blank
   \newr@f{#1}%
   \w@rnwrite{Reference #1=\number\r@fcount\space is not cited up to now.}\fi%
  \expandafter\gdef\csname r@ftext\csname r@fnum#1\endcsname\endcsname%
  {\writer@f#1>>#2#3\par}}

\def\ignoreuncited{%   redefine \refis if ignoring uncited references
   \def\refis##1##2##3\par{\ifuncit@d{##1}%
    \else\expandafter\gdef\csname r@ftext\csname r@fnum##1\endcsname\endcsname%
     {\writer@f##1>>##2##3\par}\fi}}

\def\r@ferr{\endreferences\errmessage{I was expecting to see
\noexpand\endreferences before now;  I have inserted it here.}}
\let\r@ferences=\references
\def\references{\r@ferences\def\endmode{\r@ferr\par\endgroup}}

\let\endr@ferences=\endreferences
\def\endreferences{\r@fcurr=0%		  Save old \endreferences, redefine
  {\loop\ifnum\r@fcurr<\r@fcount%	  Loop over refnum and produce text
    \advance\r@fcurr by 1\relax\expandafter\r@fis\expandafter{\number\r@fcurr}%
    \csname r@ftext\number\r@fcurr\endcsname%
  \repeat}\gdef\r@ferr{}\endr@ferences}

% Save old \endpaper, redefine it to write parting message.

\let\r@fend=\endpaper\gdef\endpaper{\ifr@ffile
\immediate\write16{Cross References written on []\jobname.REF.}\fi\r@fend}

\catcode`@=12

\citeall\refto		% These macros will generate citations
\citeall\ref		%
\citeall\Ref		%

\def\a{{\alpha}}

\def\s{\sigma}
\def\half{{1 \over 2}}
\def\ra{{\rangle}}
\def\la{{\langle}}

\def\ih{{i \over \hbar}}

\def\x{{\bar x}}

\def\D{{\cal D}}

\def\ria{{\rightarrow}}
\def\lajos{Visiting Research Fellow at: Theory Group,
Blackett Laboratory, Imperial College, London, SW7 2BZ, UK. 
Email address: diosi@rmki.kfki.hu} 
\def\jjh{{Email address: j.halliwell@ic.ac.uk}}

\centerline{\bf Coupling Classical and Quantum Variables}
\centerline{\bf using Continuous Quantum Measurement Theory}

\vskip 0.3in

\author{Lajos Di\'osi\footnote{$^{\dag}$}{\lajos} }
\affil
KFKI Research Institute for Particle and Nuclear Physics
H-1525 Budapest 114
POB 49
Hungary

\bigskip
\centerline {\rm and}

\author{Jonathan J. Halliwell\footnote{$^*$}{\jjh} }
\affil
Theory Group, Blackett Laboratory
Imperial College, London SW7 2BZ
UK
\vskip 0.2in
%\centerline{\rm PACS Numbers: 03.65.-w, 03.65.Bz, 05.40.+j, 42.50.-p}
\centerline {\rm Preprint IC 96-97/46, quant-ph/9705008. May, 1997}
\vskip 0.2in 
\centerline {\rm Submitted to {\sl Physical Review Letters}}
\vskip 0.2in
\abstract
{
We propose a system of equations to describe the interaction of a
quasiclassical variable $X$ with a set of quantum variables $x$ that
goes beyond the usual mean field approximation. The idea is to
regard the quantum system as continuously and imprecisely measured by
the classical system. The effective equations of motion for the
classical system therefore consist of treating the quantum variable
$x$ as a stochastic c-number $\x (t) $ the probability distibution
for which is given by the theory of continuous quantum measurements.
The resulting theory is similar to the usual mean field equations
(in which $x$ is replaced by its quantum expectation value) but with
two differences: a noise term, and more importantly,  the state of
the quantum subsystem  evolves according to the stochastic
non-linear Schr\"odinger equation of a continuously measured system.
In the case in which the quantum system starts out in a
superposition of well-separated localized states, the classical
system goes into a statistical mixture of trajectories, one
trajectory for each individual localized state.
} 

\endtopmatter

A variety of problems in a number of different fields involve
coupling quantum variables to variables that are effectively
classical. A case of particular interest is quantum field theory in
curved space time, where one would often like to understand how a
quantized matter field affects a classical gravitational field.
The most commonly postulated way of modeling this situation is the
semiclassical Einstein equations [\cite{Mol,Ros}]:
$$
G_{\mu \nu} = 8 \pi G \la T_{\mu \nu} \ra .
\eqno(1)
$$
Here, the left hand side is the Einstein tensor of the classical
metric field $g_{\mu \nu} $ and the right hand side is the
expectation value of the energy momentum tensor of a quantum
field. Much effort has been put into finding solutions to Eq.(1),
especially in the context of black hole physics.

Yet one cannot realistically expect that an equation such as (1)
could be valid in more than a very limited set of circumstances. One
would expect it to be valid, for example, only when the fluctuations
in energy density are small [\cite{For,KuF}], and it is not
difficult to produce situations in which its predictions are not
physically reasonable [\cite{Kib1,PaG}]. In particular, when the
quantum state of the matter field consists of a superposition of two
well-separated localized states, Eq.(1) suggests that the
gravitational field couples to the average energy density of the two
states, whilst physical intuition suggests that the gravitational
field feels the energy of one or other of the localized matter
states, with some probability. It therefore becomes of interest to
ask, is there a way of going beyond the naive mean field equations
which sensibly accommodates a wide class of non-trivial matter
states, but without having to tackle the considerably more difficult
question of quantizing the gravitational field?

In this letter we will present a simple scheme  for coupling
classical and quantum variables which goes far beyond the naive mean
field equations, and produces intuitively sensible results in the
key case of superposition states. We will not address the full
problem of the semiclassical Einstein equations (1), but rather, we
will concentrate on a simple model in which the scheme is easily
presented and perhaps verified.  Our attempt to describe the
coupling of classical and quantum variables is of course one of many
[\cite{Ale,Dio2,Zou}].

We consider a classical
particle with position $X$ in a potential $V(X)$
coupled to a 
harmonic oscillator with position $x$ which will later be quantized. 
The action is
$$
S= \int dt \ \left( \half M \dot X^2 + V(X) + \half m \dot x^2 - \half m
\omega^2 x^2 - \lambda X x \right) .
\eqno(2) 
$$ 
Hence the classical equations of motion are 
$$ 
\eqalignno{ 
M \ddot X + V'(X) + \lambda x &= 0 ,
&(3)\cr 
m \ddot x + m \omega^2 x + \lambda X &= 0 . 
&(4) \cr }
$$
The naive mean field approach involves replacing (3) with the
equation
$$
M \ddot X + V'(X) + \lambda \la \psi | \hat x | \psi \ra = 0 ,
\eqno(5)
$$
and replacing (4) with the Schr\"odinger equation 
$$
{d \over dt} | \psi \ra = - \ih \left( 
\hat H_0 + \lambda X \hat x \right) | \psi \ra 
\eqno(6)
$$
for the quantum
particle. $\hat H_0$ is the Hamiltonian of the quantum particle
(in this case a harmonic oscillator) and
$-X(t)$ is regarded as an external classical force.
As stated above, however, we
do not expect this scheme to be valid very widely.

Our new approach is to think of the  classical particle as in some
sense ``measuring'' the quantum particle's position and responding
to the measured c-number result $\x$.  (A precursor to this idea may
be found in Ref.[\cite{Dio3}]).

Consider first, therefore, the 
consequences of standard quantum measurement theory for the evolution
of the coupled classical and quantum systems over a small interval of
time $\delta t $. The state $| \psi
\ra $ of the quantum system will evolve, as a result of the measurement,
into the (unnormalized) state
$$
| \Psi_{\x} \ra = \hat P_{\x}  e^{- i \hat H \delta t} | \psi \ra
\eqno(7)
$$
where $\hat H = \hat H_0 + \lambda X \hat x $ 
and $\hat P_{\x}$ is a projection operator
which asks whether the position of the quantum particle
is $\x$, to within some precision.
The probability that the measurement yields the result
$\x$ is given by $ \la \Psi_{\x} | \Psi_{\x} \ra $. It is then
natural to suppose that the classical
particle, in responding to the measured result,
will evolve during this small time interval
according to the equation of motion
$$
M \ddot X + V'(X) + \lambda \x  = 0,
\eqno(8)
$$
with probability $ \la \Psi_{\x} | \Psi_{\x} \ra $.

Now we would like to 
repeat the process for an arbitrary number of time steps and
then take the continuum limit. If $\hat P_{\x}$ is
an exact projection operator, {\it i.e.}, one for which 
$\hat P^2_\x = \hat P_\x$, the continuum limit is trivial and 
of no interest (this is the watchdog effect). However, standard
quantum measurement theory has been generalized to a well-defined
and non-trivial
process that acts continuously in time by replacing $\hat P_{\x}$ with a
positive operator-valued measure (POVM) [\cite{BLP,CaM,Dio4,BeS,Dio3}]. 
The simplest example, which we use here, is a Gaussian,
$$
\hat P_{\x} = {1 \over ( 4 \pi \Delta^2 )^{\half} }
\ \exp \left( - { ( \hat x - \x )^2
\over 4 \Delta^2 } \right) 
\eqno(9)
$$
and the continuum limit involves taking $\Delta \ria \infty $ as
$ \delta t \ria 0 $ in such a way that $\Delta^2 \delta t $ is held
constant. 
The evolution of the wave function
of the quantum system is then conveniently expressed in terms of a
path-integral expression for the unnormalized wave function:
$$
\eqalignno{
\Psi_{[\x (t)]} (x',t') = \int \D x
& \exp \left( \ih \int_0^{t'} dt \left( \half m \dot x^2 - \half m \omega^2
x^2 - \lambda x X \right) \right)
\cr
\ \times \exp & \left( - {\lambda^2\over\hbar^2}
\int_0^{t'} dt { ( x - \x )^2 \over 4 \s^2 } \right)
\ \Psi (x_0, 0 ).
&(10) \cr}
$$
Here, the integral is over paths $ x(t) $ satisfying $x(0) = x_0 $
and $x(t') = x' $.
The classical particle at each moment of time evolves according to
Eq.(8), where the functional probability distribution of the entire
measured path $\x(t)$ takes the form:
$$
p [\x (t) ] = \la \Psi_{[ \x (t) ]} | \Psi_{[ \x (t) ]} \ra.
\eqno(11)
$$

The formula (10) contains an undetermined parameter $\s$ tuning the
precision of the measurement. (We have chosen to include the factor
of $\lambda^2$ in the parametrization of the width of the
measurement in order to conveniently accommodate the physically
expected feature that (10) must reduce to unitary evolution when
$\lambda = 0 $).
To obtain a plausible estimate as
to the value of $\s$, consider what it means  
to be classical. We take the point of view that there are
no {\it fundamentally} classical systems in the world, only quantum
systems that are effectively classical under certain conditions.
The most important condition that needs to be satisfied for a system
to be essentially classical is {\it decoherence} -- interference
between histories of certain types of variables (in this case
position) must be destroyed (see, for example, Refs.[\cite{GeH,JoZ,Zur}]).
Decoherence is typically brought about
by some kind of coarse-graining procedure. A commonly used procedure
is to couple to a heat bath and then trace it out. Whichever method
is used to produce decoherence, a generic result is that the
decohered variables are subject to fluctuations (as a result of the
interaction with the heat bath, for example). In contrast to
fundamentally classical systems, effectively classical systems
therefore always suffer a minimal amount of imprecision due to these
fluctuations. (There are also of course the ubiquitous quantum
fluctuations but these are typically much smaller). The consequence
of this is that the effectively classical system will necessarily be
limited in the precision with which it can measure the quantum
system, because of its own intrinsic imprecision. 
An estimation of $\s$ ought therefore to be possible from the
fluctuations in the classical system.

To be concrete, suppose that
in the absence of a coupling to the quantum particle, the classical
particle suffers an imprecision $ \Delta F  $ in the
degree to which the classical field equations are satisfied.
This means that the distribution of $X(t)$ is expected to be proportional
to the Gaussian functional [\cite{GeH,Hal2}],
$$
\exp \left( - { 1 \over 2 ( \Delta F )^2 } \int dt \left(
M \ddot X + V'(X) \right)^2 \right).
\eqno(12)
$$
An example is the case of a thermal environment
producing the decoherence and fluctuations in $X$, in which case
$(\Delta F)^2 $ is of order $ M \gamma k_B T $, where $\gamma $ is the
dissipation of the environment and $T$ its temperature. 
On dimensional grounds, given the coupling between $X$ and $x$ in
(2), a reasonable choice for $\s$ is $ \s \sim \Delta F /  \hbar $,
(so $\s^2 \sim M \gamma k_B T /\hbar^2$ in the case of thermal
fluctuations). We will see further evidence for this choice below.

The scheme is therefore as follows. We solve the equations (8) and (10)
where $\x (t) $ is regarded as a stochastic variable whose
probability distribution is given by (11). The final result is
therefore an ensemble of $\x$-dependent
classical and quantum trajectories respectively for the 
%large and respectively for the small 
two
particles, with an interdependent probability distribution. 

It turns out that this system (8), (10), (11) can be rewritten in such a
way that brings it closer to the form of the naive mean field
equations (5), (6). The basic issue is that Eq.(11) gives the probability
for an entire history of measured alternatives, $\x (t)$. Yet the
naive mean field equations (5), (6) are evolution
equations defined at each moment of time. This therefore leads one
to ask, is it possible to rewrite the system (8), (10), (11) in terms of
evolution equations?

This is indeed possible. Consider the basic process (7)
with the Gaussian projector (9), but in addition let the 
the state vector be normalized at each time step. 
Then denoting the normalized state at
each time by $ | \psi \ra $, and taking the continuum limit in the
manner indicate above,
it is readily shown [\cite{Dio4}] that $ | \psi \ra $ obeys
a stochastic non-linear equation describing a system undergoing
continuous measurement:
$$
{d \over dt} | \psi \ra = \left( - \ih ( \hat H_0 + \lambda X \hat x ) 
- {\lambda^2 \over  4 \hbar^2 \s^2} 
( \hat x - \la \hat x \ra )^2 \right) | \psi \ra 
+ {\lambda \over 2 \hbar \s } 
\left( \hat x - \la \hat x \ra \right) | \psi \ra \eta (t).
\eqno(13)
$$
Here, $\eta (t)$ is the standard Gaussian white noise, 
with linear and quadratic means,
$$ 
\la \eta (t) \ra_S = 0, \quad \quad
\la \eta (t) \eta (t') \ra_S = \delta (t-t').
\eqno(14)
$$
where $ \la \ \ra_S $ denotes stochastic averaging.
The noise terms are to be interpreted in the sense of Ito.
The measured value $\x$ is then related to $\eta $ by
$$
\x = \la \psi | x | \psi \ra + { \hbar \s \over \lambda} \eta (t). 
\eqno(15)
$$

Hence, the final equations that replace (5), are
$$
M \ddot X + V'(X) + \lambda \la \psi | 
\hat x | \psi \ra + \hbar \s \eta (t)  = 0
\eqno(16)
$$
and (6) is replaced by the stochastic non-linear equation (13).
Referring back to our explanation of classicality,
we see that, indeed, the classical particle suffers a 
random force independent of its coupling to the 
quantum particle, leading formally to the
distribution (12). In the case of a thermal environment, 
the random force should be $\sqrt{2 M \gamma k_B T}\eta(t)$,
in order to coincide with the standard Langevin equation of
classical Brownian motion, and this is indeed the case (if 
the numerical factor in the choice of $\s$ is chosen so that
$\s^2 = 2 M \gamma k_B T / \hbar^2 $).

There are two differences between the system, (13)--(16) and the
naive mean field equations. One is the noise term, $\eta$. In
Eq.(16) the noise clearly describes  fluctuations about the naive
semiclassical trajectories. This sort of modification to the
semiclassical Einstein equations has been considered  previously
[\cite{KuF,HuM}].

More important is the novelty that the state $|\psi \ra $ evolves according
to the stochastic non-linear equation (13), and hence its evolution
is very different to that under the usual Schr\"odinger equation.
In particular, it may be shown that all solutions to (13) undergo {\it
localization} [\cite{Dio5,GP1,SaG,Per,HZ1,HZ2}] on a time scale which
might be extremely short compared to the oscillator's freqency $\omega$.
That is, every initial state rapidly evolves to a generalized
coherent state centred around values $ \la \hat x \ra $, 
$ \la \hat p \ra $
undergoing classical Brownian motion. (The results cited above are
readily extended to the case here in which the Hamiltonian
contains a linear coupling to an external force $-X(t)$).
Which particular solution the state becomes centred around depends 
statistically on the initial state of the system.
For an initial state consisting of a superposition of
well-separated coherent states,
$$
| \psi \ra = \a_1 | x_1 p_1 \ra + \a_2 | x_2 p_2 \ra
\eqno(17)
$$
the state after localization time will, with probability $| \a_1 |^2 $,
be as if the initial state were just $ | x_1 p_1 \ra $, and
with probability $| \a_2 |^2 $, will be as if the initial state were
just $ |x_2 p_2 \ra $ [\cite{HZ1,HZ2}].
The localization time $\sim 1/\s^2(x_1-x_2)^2$ 
becomes, with our previous choice $\s^2 \sim  M \gamma k_B T /\hbar^2$, 
very short indeed if the classical particle has a large 
mass $M$.

Hence in the new semiclassical equations
(13)--(16), effectively what happens is that 
we solve separately for the two initial states $| x_1 p_1 \ra $
and $| x_2 p_2 \ra $, and the 
%large 
 classical 
particle then follows the first
solution with probability $ | \a_1 |^2 $ and the second with
probability $ | \a_2 |^2 $.

In simple terms, therefore, an almost classical system interacting
through position with a  quantum system in a superposition state
(17), ``sees'' one or other of the superposition states, with some
probability, and not the mean position of the entire state. This is
the key case for which the naive mean field equations fail to give
intuitively sensible results [\cite{PaG,Dio}]. 

It is interesting to note that non-linear Schr\"odinger equations
have been considered before in the context of the semiclassical
Einstein equations [\cite{Kib1,Kib2,Mie}], because the combined 
system consiting of (1) together with the Schr\"odinger equation for
the quantum state is non-linear. The motivation here is rather
different. The equation (13) used here arises because it gives a
phenomenological description of continuous measurement. 

Similar results are obtained with different types of couplings, for
example to momentum or to energy [\cite{Hal3}]. Obviously an
important challenge is to extend to quantum field theories and hence
to obtain a generalization of Eq.(1). This would mean confronting
the difficult issues of covariance and non-renormalizability. This
will be discussed elsewhere.

We have presented a scheme for coupling classical and quantum
variables which appears to be reasonable on physical grounds and
give intuitively sensible results. It is based on the premise that
the interaction between the classical and quantum variables may be
regarded as a quantum measurement. The mathematics of continuous
quantum measurement theory then fixes the overall structure of the
scheme, but an additional physical argument is required to fix the
parameter describing the precision of the measurement.

In the simple model we considered here, the quantum theory of the
whole closed system (including the possible environmental degrees of
freedom) exists.  In this case it is therefore reasonable to explore
the possibility that the scheme presented here emerges as an 
effective theory  under suitable conditions (taking into account the
requirements for the classicality above). In a longer more detailed
paper it will be shown, using the decoherent histories approach to
quantum theory, how an effective theory very similar to this scheme
may arise [\cite{Hal3}]. This confirms the physical arguments given
above for choice of the width $\s$ of the continuous measurements.
Furthermore, the theory of continuous quantum measurements is in
fact closely related to the so-called hybrid representation of composite
quantum systems  [\cite{Dio2,Dio6,Dio7}], thus providing an alternative 
framework for examining the emergence of the scheme. 

We have presented the scheme here without appealing to more
fundamental origins because we believe that it stands on its own
terms as a simple and plausible phenomenological model.
Furthermore, since the underlying quantum theory of the classical
system may not be known, it is important to understand how such
phenomenological models are constructed directly.  We do not claim,
however, that our scheme eliminates all known controversies of the naive
mean field method or of non-linear quantum theories generally
[\cite{Gis}].

\head{\bf Acknowledgements}

L.D. was supported by a Visiting Research Fellowship from EPSRC and
by grant number OTKA T016047. We are grateful to Chris Isham for
useful comments.

\references

\def\pr{{\sl Phys. Rev.\ }}
\def\prl{{\sl Phys. Rev. Lett.\ }}

\def\cmp{{\sl Comm. Math. Phys.\ }}
\def\np{{\sl Nucl. Phys.\ }}

\def\annp{{\sl Ann. Phys. (N.Y.)\ }}

\refis{Ale} I.V.Aleksandrov, Z.Naturf. {\bf 36A}, 902 (1981);
A.Anderson, \prl {\bf 74}, 621 (1995);
% QUANTUM BACKREACTION ON CLASSICAL VARIABLES
\prl {\bf 76}, 4090 (1996);
% CAN CLASSICAL AND QUANTUM VARIABLES HAVE A CONSISTENT MUTUAL
%      INTERACTION - REPLY
W.Boucher and J.Traschen, \pr {\bf D37}, 3522 (1988);
%SEMICLASSICAL PHYSICS AND QUANTUM FLUCTUATIONS      
K.R.W.Jones, \prl {\bf 76}, 4087 (1996);
L.Di\'osi, \prl {\bf 76}, 4088 (1996);
I.R.Senitzky, \prl {\bf 76}, 4089 (1996).
 
\refis{BLP} A.Barchielli, L.Lanz and G.M.Prosperi,
{\sl Il Nuovo Cimento} {\bf 72B}, 79 (1982).
% A model for macroscopic description and continual observations 
% in quantum mechanics.

\refis{BeS} V.P.Belavkin and P.Staszewski, Phys.Rev. {\bf A45}, 1347 (1992).

\refis{CaM} C.M.Caves and G.J.Milburn, {\sl Phys.Rev.} {\bf A36}, 5543 (1987).

\refis{Dio} L.Di\'osi, {\sl Phys.Lett.} {\bf 105A}, 199 (1984).

\refis{Dio2} L.Di\'osi, ``A True Equation to Couple Classical and
Quantum Variables'', preprint quant-ph/9510028 (1995),

\refis{Dio3} L.Di\'osi, {\sl Phys.Rev.} {\bf A42}, 5086 (1990).

\refis{Dio4} L.Di\'osi, {\sl Phys.Lett.} {\bf 129A}, 419 (1988).
% Continuous quantum measurement and Ito formalism

\refis{Dio5} L.Di\'osi, {\sl Phys.Lett} {\bf 132A}, 233 (1988).
% Localized solutions of a simple nonlinear quantum langevin equation.

\refis{Dio6} L.Di\'osi, {\sl Quantum Semiclass.Opt.} {\bf 8}, 309
(1996).
% Exact semiclassical wave equation for stochastic quantum optics.

\refis{Dio7} L.Di\'osi, preprint quant-ph/9610037 (1996). To appear
in {\it Fundamental Problems in Quantum Physics}, edited by
M.Ferrero and A. van der Merwe (Kluwer, Denver, 1997).

\refis{For} L.H.Ford, \annp {\bf 144}, 238 (1982).

\refis{GeH} M. Gell-Mann and J. B. Hartle, 
%in {\it Complexity, Entropy 
%and the Physics of Information, SFI Studies in the Sciences of Complexity},
%Vol. VIII, W. Zurek (ed.) (Addison Wesley, Reading, 1990); and in
%{\it Proceedings of the Third International Symposium on the Foundations of
%Quantum Mechanics in the Light of New Technology}, S. Kobayashi, H. Ezawa,
%Y. Murayama and S. Nomura (eds.) (Physical Society of Japan, Tokyo, 1990);
{\sl Phys.Rev.} {\bf D47}, 3345 (1993).

\refis{Gis} N.Gisin, {\sl Helv.Phys.Acta} {\bf 62}, 363 (1989).
% Stochastic quntum dynamics and relativity.

\refis{GP1} N. Gisin and I.C.Percival, {\sl J.Phys.} {\bf A26},
2233 (1993); {\bf A26}, 2245 (1993).

%\refis{Gri} R.Griffiths, {\sl J.Stat.Phys.} {\bf 36}, 219 (1984).

\refis{Hal2} J.J.Halliwell, {\sl Phys.Rev.} {\bf D48}, 4785 (1993).
% Quantum-Mechanical Histories and the Uncertainty 
% Principle. II. Fluctuations about Classical Predictability.

\refis{Hal3} J.J.Halliwell, ``Effective Theories of Coupled
Classical and Quantum Variables from Decoherent Histories: A New
Approach to the Backreaction Problem'', Imperial College preprint
96-97/45, quant-ph/9705005 (1997)

\refis{HZ1} J.J.Halliwell and A.Zoupas,
{\sl Phys.Rev.} {\bf D52}, 7294 (1995). 
% ``Quantum State Diffusion, Density Matrix Diagonalization
% and Decoherent Histories: A Model''.

\refis{HZ2} J.J.Halliwell and A.Zoupas, 
``Post-decoherence density matrix propagator for quantum 
Brownian motion'',
IC preprint 95-96/67, quant-ph/9608046 (1996).
Accepted for publication in {\sl Phys.Rev.D} (1997).

%\refis{HaH} J.B.Hartle and G.T.Horowitz, \pr {\bf 24}, 257 (1981).
% Ground state expectation value of the metric in the 1/N
% or semiclassical approximation to quantum gravity.

\refis{HuM} B.L.Hu and A.Matacz, \pr {\bf D51}, 1577 (1995).
% Backreaction in semiclassical gravity: The Einstein-Langevin equation.

\refis{JoZ} E.Joos and H.D.Zeh, {\sl Zeit.Phys.} {\bf B59}, 223
(1985).
%{\it The Emergence of Classical Properties through Interaction with
%an Environment.}

\refis{Kib1} T.W.B.Kibble, in {\it Quantum Gravity 2: A Second Oxford
Symposium}, edited by C.J.Isham, R.Penrose and D.W.Sciama (Oxford
University Press, New York, 1981).

\refis{Kib2} T.W.B.Kibble, \cmp {\bf 64}, 73 (1978);
T.W.B.Kibble and S.Randjbar-Daemi, {\sl J.Phys.} {\bf A13},
141 (1980).

\refis{KuF} C-I.Kuo and L.H.Ford, \pr {\bf D47}, 4510 (1993).
% Semiclassical gravity theory and quantum fluctuations.

\refis{Mie} B.Mielnik, \cmp {\bf 37}, 221 (1974).

\refis{Mol} C.Moller, in {\it Les Theories Relativistes de la
Gravitation}, edited by A.Lichnerowicz and M.A.Tonnelat (CNRS,
Paris, 1962).

%\refis{Omn} R.Omn\`es, {\it The Interpretation of Quantum Mechanics}
%(Princeton University Press, Princeton, 1994);
%{\sl Rev.Mod.Phys.} {\bf 64}, 339 (1992), and references therein.

\refis{PaG} D.N.Page and C.D.Geilker, \prl {\bf 47}, 979 (1981).
% Indirect evidence for quantum gravity.

\refis{Per} I.C.Percival,  {\sl J.Phys.} {\bf A27}, 1003 (1994).

\refis{Ros} L.Rosenfeld, \np {\bf 40}, 353 (1963).

\refis{SaG} Y.Salama and N.Gisin, {\sl Phys. Lett.} {\bf 181A}, 269, (1993).

\refis{Zou} A.Zoupas, ``Coupling of Quantum to Classical in the
Presence of a Decohering Environment'', Imperial College preprint (1997).

\refis{Zur} W. Zurek, {\sl Prog.Theor.Phys.} {\bf 89}, 281 (1993);
{\sl Physics Today} {\bf 40}, 36 (1991);
in, {\it Physical Origins of Time Asymmetry}, edited by 
J.J.Halliwell, J.Perez-Mercader and W.Zurek (Cambridge
University Press, Cambridge, 1994).

\endreferences

\end